\documentstyle[epsfig,amsmath,amssymb,bbold]{article}

\def\slash#1{\setbox0=\hbox{$#1$}               
   \dimen0=\wd0                                 
   \setbox1=\hbox{/} \dimen1=\wd1               
   \ifdim\dimen0>\dimen1                        
      \rlap{\hbox to \dimen0{\hfil/\hfil}}      
      #1                                        
   \else                                        
      \rlap{\hbox to \dimen1{\hfil$#1$\hfil}}   
      /                                         
   \fi}                                         %

\def\be{\begin{eqnarray}}
\def\ee{\end{eqnarray}}

\newcommand{\bea}{\begin{eqnarray}}
\newcommand{\eea}{\end{eqnarray}}       

\begin{document}


\title{Renormalization Group Flow in large $N_c$}
\author{Jochen Meyer$^{a}$, Kai Schwenzer$^{b}$, Hans--J\"urgen
  Pirner$^{c}$ \\[2mm]
\small{Institut f\"ur Theoretische Physik, Universit\"at Heidelberg,} \\
\small{Philosophenweg 19, 69120 Heidelberg, Germany} \\[4mm]
Aldo Deandrea$^{d}$ \\[2mm]
\small{Institut de Physique Nucl\'eaire, B\^atiment Paul Dirac,} \\
\small{ Universit\'e Claude Bernard Lyon I,} \\ 
\small{43, bd du 11 Novembre 1918, 69622 Villeurbanne Cedex, France} \\[4mm]
\small{(a) jochen@tphys.uni-heidelberg.de, (b)
  kai@tphys.uni-heidelberg.de,} \\ 
\small{(c) pir@tphys.uni-heidelberg.de, (d)
  deandrea@ipnl.in2p3.fr}}

\maketitle

\begin{abstract}
  We calculate renormalization group flow equations for the linear
  $\sigma$-model in large $N_c$ approximation. The flow equations
  decouple and can be solved analytically. The solution is
  {\em equal} to a self consistent solution of the NJL model in the same
  approximation, which shows that flow equations are a promising
  method to extend the calculation to higher order in $1/N_c$. 
  Including explicit chiral symmetry
  breaking, the large $N_c$ approximation describes physics reasonably
  well. We further compare the analytic solution to the usually used
  polynomial truncation and find consistency.
\end{abstract}


\newpage
\section{RG flow equations of the linear $\sigma$-model}
The physics of strong interactions in the nonperturbative strong
coupling regime cannot be obtained directly from QCD. Especially, it
is not yet possible to integrate out the gluonic degrees of freedom in a
rigorous way \cite{Meggiolaro:2000kp}. From 1-gluon exchange or
instanton models one is lead to a local 4-fermion interaction
described by the Nambu$-$Jona-Lasinio (NJL) model. The NJL model is
non-renormalizable and therefore includes an explicit cutoff scale. It
is commonly considered in large $N_c$ approximation
\cite{Klevansky:1992qe}. If fluctuations in the quark condensate are
taken into account one needs an additional cutoff \cite{Oertel:2000fk}, which
reduces the predictive power of the model. Such an additional cutoff
is not needed in the renormalization group (RG) framework, when the
NJL-model is embedded into the linear $\sigma$-model. The main
aim of this work is to clarify the connection between the NJL model
and the RG flow equations for the associated linear
$\sigma$-model in the large $N_c$ approximation. \\
RG flow equations describe the average of an effective action and
represent the continuum analogue of a block spin
transformation. Various kinds of exact flow equations have proved
successful in describing a huge set of different systems \cite{RGbasics}. For
an introduction to the RG method we refer to the review \cite{RGreview}.  
The flow equations are ultraviolet (UV) and infrared (IR) finite
through the introduction of a scale $(k)$ - dependent cutoff
function. Degrees of freedom with momenta larger than the
coarse--graining scale $k$ are integrated out. The
solution of the flow equations includes summations of different diagrams
and is similar to the solution of a coupled set of Schwinger Dyson
equations. We use the Schwinger proper time regularization method
\cite{Schaefer:1999em}, which preserves all symmetries of
the theory and keeps the physical interpretation of the evolution
equations particularly simple for phenomenological
applications. This simplicity makes these RG improved 1-loop equations
interesting, although they are not exact RG flow equations
\cite{Litim:2001ky}. \\
The lagrangian of the NJL model with an explicit current quark mass $m_c$
reads in Euclidian spacetime
\begin{equation}
\label{eq:lagrangian-njl}
{\cal L}_{\mbox{\tiny NJL}} = \bar{q} ({\slash \partial}+m_c) q 
- G \left( (\bar{q} q)^2 + (\bar{q} 
i\vec{\tau}\gamma_5 q)^2 \right) \; .
\end{equation}
A partial bosonization by the introduction of auxiliary scalar fields
$\sigma$ and $\pi$ gives the associated linear $\sigma$-model \cite{Ripka}
\begin{equation}
\label{eq:lagrangian-initial}
{\cal L}_{\mbox{\tiny uv}} = \bar{q} {\slash \partial} q 
+ g \, \bar{q} \left( \sigma  + i\vec{\tau}\vec{\pi} \gamma_5 \right) q 
+ \frac{m^2}{2} \left( \sigma^2+\vec \pi^2 \right) - \delta \, \sigma \; .
\end{equation}
The respective couplings are connected by 
\begin{equation}
2 G = \frac{g^2}{m^2} \quad , \quad m_c=\frac{g \, \delta}{m^2} \; .
\label{eq:gnjl}
\end{equation} 
In order to solve the theory it is sufficient to compute the effective
action, which is the generating functional of one particle irreducible
Feynman diagrams. It involves in principle arbitrary higher order
terms consistent with the symmetry of the initial action.  Using RG
methods we are going to calculate the effective action in a suitable
truncation.  Surely we cannot account for all possible terms, so our
approximation is to assume that the linear $\sigma$-model with quarks
is a valid description of nature at a scale of $0.2$ GeV$\lesssim\! k
\lesssim 1$ GeV. In this region the gluon degrees of freedom are
supposed to be already frozen out, but confinement is not yet
relevant. It remains to be investigated whether the effects of other $
\bar q q $-resonances are sufficiently taken into account by quarks
coupling to $(\sigma, \vec \pi)$ fields.
The physics of the model is governed by chiral symmetry, which is
spontaneously broken in the IR. The Euclidian UV action of the general
linear $\sigma$-model is given by
\begin{equation}
  \label{eq:uv-action}
  S[\Phi,\bar q, q] = \int \! d^4x \left( Z_q \, \bar{q} {\slash
  \partial} q +\frac{1}{2} Z_\Phi \left( \partial_\mu \Phi \right)
  \left( \partial^\mu \Phi \right) + g \, \bar{q} M q + U
  \right) \; .
\end{equation}
The quark- and meson fields $\bar q, q, \Phi, M$ are bare
fields, where $\Phi\!=\!(\sigma,\vec \pi)$ is the $O(4)$-
and $M\!=\!\sigma\!+\!i\vec\tau \vec\pi \gamma_5$ the chiral $SU(2)_L \otimes
SU(2)_R$-representation of the meson fields. We consider this
action to be preserved during the evolution and choose
the starting values 
\begin{align}
 Z_\Phi &= 0 \; , \\ 
 Z_q &= 1 \; ,
\end{align}
and the UV potential 
\begin{equation}
\label{eq:uv-potential}
U_{\mathrm{uv}}=\frac{m^2}{2} \, \Phi^2 -\delta \, \sigma \; , 
\end{equation}
in order to obtain Eq. (\ref{eq:lagrangian-initial}) as UV lagrangian.
For comparison with the standard NJL approximation we will work in the
following in the chiral limit $m_c\!=\!\delta\!=\!0$, but since the
term linear in $\sigma$ does not get renormalized
\cite{Zinn-Justin}, the following computation is also valid for
explicit symmetry breaking. This case will be discussed in Sec. 3.
The effective action $\Gamma$, which depends only on averaged fields,
can be computed by doing a saddle point expansion corresponding
to a one loop approximation (cf. \cite{Zinn-Justin}):
\begin{align} 
  \label{eq:effective-action}  
  &\Gamma[\Phi,\bar q, q] = \, S[\Phi,\bar q, q] - \frac{1}{2}
      \mathrm{Tr} \, \log \left( \frac{\delta^2 S[\Phi,\bar q,
      q]}{\delta \bar q(x) \delta q(y)} \right) \\
  &+\frac{1}{2} \mathrm{Tr} \, \log \left(
  \frac{\delta^2 S[\Phi,\bar q, q]}{\delta \Phi^i(x) \delta
      \Phi^j(y)} - 2 \, \frac{\delta^2 S[\Phi,\bar q,
      q]}{\delta \Phi^i(x) \delta q(y)}
  \left( \frac{\delta^2 S[\Phi,\bar q, q]}{\delta \bar q(x) \delta
  q(y)} \right)^{-1} \frac{\delta^2 S[\Phi,\bar q,
      q]}{\delta \bar q(x) \delta \Phi^j(y)} \right) \; . \nonumber
\end{align}
The first logarithm results from fermion loop fluctuations, whereas
the two terms in the second logarithm are the contributions of the
bosonic and the mixed loop respectively. \\
The logarithms are transformed into Schwinger proper time integrals
regularized by a heat kernel cutoff $f(k^2 \tau)$ (cf. \cite{Schaefer:1999em})
\begin{equation}
  \mathrm{log} (A)= -\int_{\frac{1}{\Lambda^2}}^\infty \! \frac{d
  \tau}{\tau} \,
  e^{-\tau A} f(k^2 \tau) \; .
\end{equation}
As cutoff the following set of functions is used
\begin{equation}
  \label{eq:co}
  f^{(n)}(k^2\tau)=\sum_{i=0}^n \frac{(k^2 \tau)^i}{i!} \, \exp(-k^2 \tau) \; .
\end{equation}
The heat kernel cutoff functions $f^{(n)}$ suppress fluctuations with
momenta below the cutoff scale $k$. Going to $k\rightarrow 0$ means to
include more and more IR modes. Finally all modes are included,
because of the limiting behaviour $f^{(n)} \rightarrow 1$. Due to the
extra $\tau$ terms in front of the exponential the
integral diverges in the UV region only through an additional
term $\sim \log \Lambda^2$. To keep the following computation simple,
we will consider the cutoff $f^{(2)}$ used in \cite{Schaefer:1999em}. It has
been shown for the case of a $O(N)$-model, that the RG equation
derived with the cutoff $f^{(2)}$ is equivalent to an exact RG
equation \cite{Litim}. \\ 
The detailed computation of the effective action is based on a
derivative expansion and will be presented in a separate paper
\cite{fullflow}. The purely bosonic part was
investigated in \cite{Bohr}. \\
The derivative of the regularized effective action with respect
to the cutoff scale $k$ yields a flow equation for the effective
action.  Identifying terms in $\frac{\partial \Gamma}{\partial k}$
with similar terms in $ S[\Phi,\bar q, q]$ we obtain
evolution equations for the effective potential $U$, coupling
constant $g$ and the wave function renormalization factors $Z_q$ and
$Z_\Phi$. \\
To improve the one loop equation
we substitute  the masses and couplings of the classical action by
the running masses and running couplings of the {\em effective action}
$\Gamma$.  This replacement turns the one loop equations into RG
improved flow equations, which include higher
loop terms successively into the proper time integral. Therefore the
flow equations can treat nonperturbative physics in the strong coupling
region \cite{RGreview}. \\
We find the following coupled set of equations for the parameters of
the model\footnote{Contrary to the equations for $U(\Phi^2)$ and
  $Z_\Phi(\Phi^2)$, which are valid for general $\Phi^2$-dependent
  functions, the given equations for the Yukawa coupling and the quark
  wave function renormalization factor are only valid for
  $\Phi$-independent $g$ and $Z_q$. The general expressions
  get correction terms which do not contribute in large $N_c$.}
\begin{align}
\label{eq:flow-eq-potential}
  k\frac{\partial U}{\partial k}
  =& -\frac{N_f N_c}{8\pi^2} \, \frac{k^6}{k^2\!+\!M_q^2}
  +\frac{1}{32\pi^2} \left( \frac{k^6}{k^2\!+\!M_\sigma^2} +3 \,
  \frac{k^6}{k^2\!+\!M_\pi^2} \right) \; ,\\ 
\label{eq:flow-g}
  k\frac{\partial g}{\partial k} 
  =& \frac{1}{16\pi^2} \, \frac{g^3}{Z_q^2 Z_\Phi} 
  \left(
    \frac{k^6 \, (2k^2\!+\!M_q^2\!+\!M_\sigma^2)}{(k^2\!+\!M_q^2)^2 \,
  (k^2\!+\!M_\sigma^2)^2} \right. \nonumber \\
  &\qquad \qquad \qquad \qquad \; \; \; \left. -3 \, \frac{k^6 \,
  (2k^2\!+\!M_q^2\!+\!M_\pi^2)}{(k^2\!+\!M_q^2)^2 \, (k^2\!+\!M_\pi^2)^2}
  \right) \; ,
\end{align}
\begin{align}
\label{eq:flow-zq}
  k\frac{\partial Z_q}{\partial k} 
  =& - \frac{1}{16\pi^2} \, \frac{g^2}{Z_q Z_\Phi}
  \left( \frac{k^6 \, (2k^2\!+\!M_q^2\!+\!M_\sigma^2)}{(k^2\!+\!M_q^2)^2
  \, (k^2\!+\! M_\sigma^2)^2} \right. \nonumber \\
  &\qquad \qquad \qquad \qquad \qquad \left. +3 \, \frac{k^6 \,
  (2k^2\!+\!M_q^2\!+\!M_\pi^2)}{(k^2\!+\! M_q^2)^2 \,
  (k^2\!+\!M_\pi^2)^2} \right) \; , \\
\label{eq:flow-zphi}
  k\frac{\partial Z_\Phi}{\partial k} =& -\frac{N_f N_c}{4\pi^2} \,
  \frac{g^2}{Z_q^2} \, \frac{k^6}{(k^2\!+\!M_q^2)^3}
  - \frac{1}{8\pi^2} \, Z_\Phi \Lambda \, \frac{k^6 \,
  (M_\sigma^2\!-\!M_\pi^2)}{(k^2 \!+\! M_\sigma^2)^2 \, (k^2 \!+\!
  M_\pi^2)^2} \; ,
\end{align}
where the effective masses are given by
\begin{align}
\label{eq:masses}
\begin{split}
M_q^2(\Phi^2) =& \frac{g^2}{Z_q^2} \, \Phi^2 \; ,\\
M_\sigma^2(\Phi^2) =&  \frac{2}{Z_\Phi} \left(\frac{\delta
  U(\Phi^2)}{\delta \Phi^2} + 2 \, \Phi^2 \, \frac{\delta^2
  U(\Phi^2)}{\left( \delta \Phi^2 \right)^2} \right) \; ,\\
M_\pi^2(\Phi^2) =&  \frac{2}{Z_\Phi} \, \frac{\delta
  U(\Phi^2)}{\delta \Phi^2} \; ,
\end{split}
\end{align} 
and the effective 4-meson coupling reads
\begin{equation}
\label{eq:4-meson-coupling}
  \Lambda(\Phi^2)=\frac{2}{Z_\Phi^2} \frac{\delta^2 U(\Phi^2)}{(\delta
  \Phi^2)^2} \; .
\end{equation}
In this set of equations all the masses and couplings are
functions of the scale $k$. The heat kernel method
gives these equations in a very simple form. Since the IR parameter
$k$ acts like a mass cutoff, the propagator structure of the
different loop contributions is still recognizable. \\
Although the effective masses and the effective 4-meson coupling in Eqs.
(\ref{eq:flow-eq-potential}-\ref{eq:flow-zphi}) are fully renormalized
quantities, the potential $U$ and the Yukawa coupling $g$ are still
bare ones. In order to obtain physical couplings one has to
express the theory by renormalized fields $q_R$, $\bar q_R$ and
$\Phi_R$. This is done by rescaling the fields to absorb the
wave function renormalization factors in front of the kinetic terms:
\begin{equation}
  \Phi_R=(Z_\Phi(\Phi_0^2))^\frac{1}{2} \, \Phi \quad , \quad
  q_R=Z_q^\frac{1}{2} \,  q \, .
\end{equation}
where $\Phi_0$ denotes the vacuum expectation value of the field $\Phi$.
The renormalized potential is a function of the renormalized meson field
$U_R(\Phi_R^2)=U((Z_\Phi(\Phi_0^2))^{-1} \, \Phi_R^2)$ and the
renormalized Yukawa coupling is related to the bare coupling as $g_R =
Z_q^{-1} \, (Z_\Phi(\Phi_0^2))^{-\frac{1}{2}} \, g$. Finally, the
physical masses and the physical 4-meson coupling are given at the
vacuum expectation value of the meson field
\begin{equation}
\label{eq:physmasses}
  m_i=M_i(\Phi_0^2) \quad , \quad i \in \{\sigma,\pi,q \} \quad ;
  \quad \lambda=\Lambda(\Phi_0^2) \; .
\end{equation}
%
%

\newpage
\section{RG in large $N_c$ and NJL gap-equation}
\begin{figure}[h]
\begin{center}
\begin{raggedright}
\epsfig{file=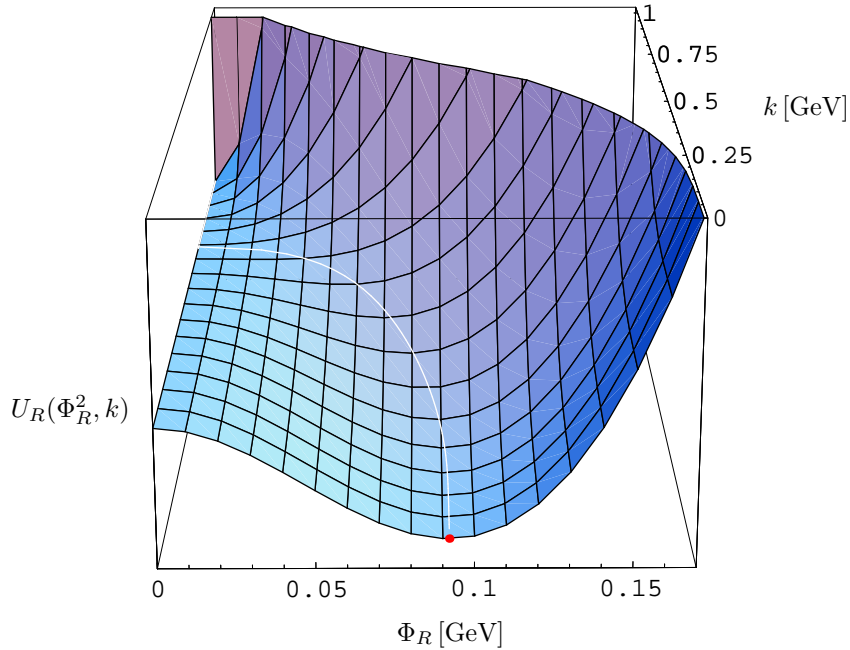}
\end{raggedright} 
\centerline{$\Phi_R \, [\mbox{GeV}]$}
\flushleft \vspace{-7.7cm} \hspace{10.1cm} {$k \, [\mbox{GeV}]$}

\vspace{3.6cm} \hspace{0.1cm} {$U_R(\Phi_R^2,k)$}
\vspace{+3.0cm} 
\caption{\label{fig:pot-plot} The evolution for the
  renormalized meson potential in large $N_c$ approximation. In the UV
  the renormalized potential is infinite for all $\Phi_R$ due to the
  vanishing wave function renormalization constant. The quadratic
  potential, which arises when $Z_\Phi$ gets finite, evolves to a
  potential with broken symmetry in the IR. The white curve shows the
  evolution of the vacuum expectation value, which corresponds to
  $f_\pi\!= \!92$ MeV at $k\!=\!0$, denoted by the endpoint.}
\end{center}
\end{figure}

\begin{figure}[h]
\begin{center}
\begin{raggedright}
\epsfig{file=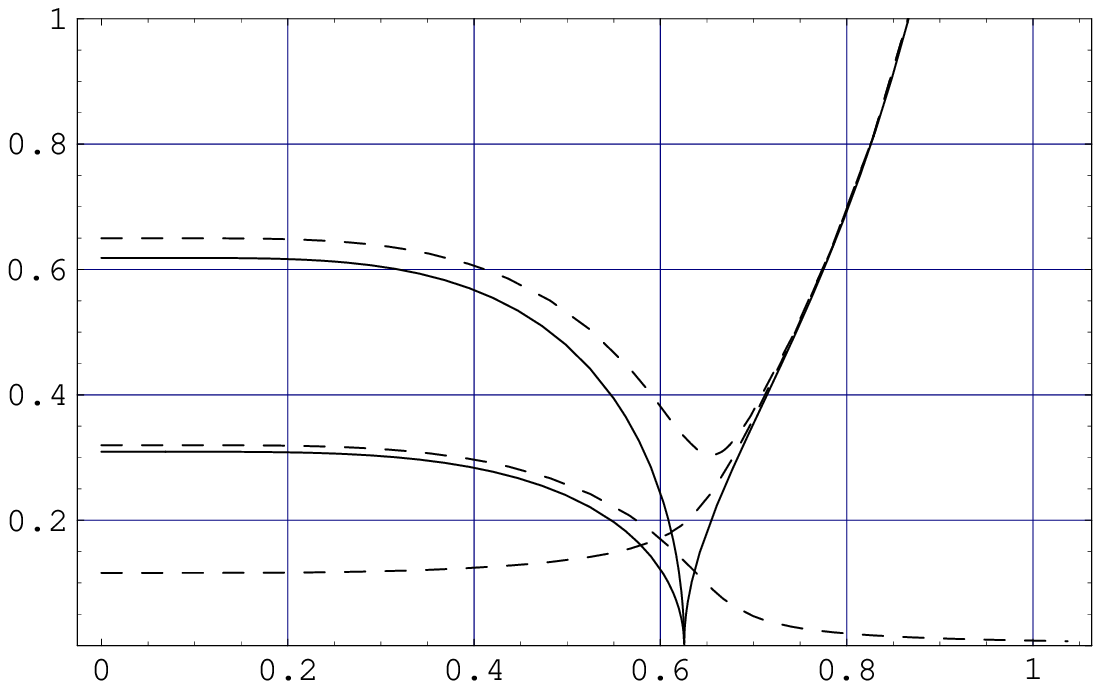,width=10cm}
\end{raggedright} 
\centerline{$k \; [\mbox{GeV}]$}
\flushleft \vspace{-1.9cm} \hspace{2.5cm} {$m_\pi$} \hspace{6.0cm} {$m_q$}

\vspace{-1.5cm} \hspace{2.5cm} {$m_q$}

\vspace{-3.0cm} \hspace{2.5cm} {$m_\sigma$} \hspace{5.6cm}
{$m_\pi\!=\!m_\sigma$}
\flushleft \vspace{0.5cm} \hspace{0.1cm} {$[\mbox{GeV}]$}
\vspace{+3.7cm} 
\caption{\label{fig:masses}The renormalized masses of the exact solution as
  functions of the scale $k$ in the chiral
  limit (solid) and with an explicit symmetry breaking due to a
  current quark mass of $7$ MeV (dashed). The quark mass in
  the symmetric regime and the pion mass in the broken regime vanish
  in the chiral limit, whereas the pion acquires a mass of
  $m_\pi\!=\!116$ MeV in the case of a finite current quark
  mass.}    
\end{center}
\end{figure}

In the large $N_c$ limit only the fermion loop is relevant.
Since the NJL effective action involves a trace over quark fields it is
proportional to $N_c$ and therefore the NJL coupling $G$ is
proportional to $1/N_c$. This also has to hold in the bosonized case
Eq. (\ref{eq:uv-action}). 
Assuming that the meson masses are independent of $N_c$, one obtains
from Eq. (\ref{eq:gnjl}) that $g$ behaves $\propto 1/\sqrt{N_c}$, in
the same way as the QCD coupling constant. This implies that the meson
fields $\Phi$ scale as ${\cal O}(\sqrt{N_c})$, which in turn means
that the effective 4-meson coupling $\lambda$ should be
proportional to $1/N_c$. \\
Due to the large $N_c$ behaviour, the quark contribution in the flow
equation of the potential $U$ Eq. (\ref{eq:flow-eq-potential}) is of
order $N_c$, whereas the meson loop term is of order one and can be
neglected. Also, for $Z_\Phi$ Eq. (\ref{eq:flow-zphi}) the second term
can be dropped since it is of order $1/N_c$ compared to the first one
which is of order one.  The wave function renormalization parameter
$Z_q$ Eq.  (\ref{eq:flow-zq}) and the bare Yukawa-coupling $g$ Eq.
(\ref{eq:flow-g}) do not evolve, since the right hand sides of the
evolution equations are of higher order in $1/N_c$ than the left hand
sides, which reflects that the corresponding fluctuations are
given by mixed loops. \\
The large $N_c$ flow equations read ($Z_q\!=\!g\!=\!1$)
\begin{align}
\label{eq:large-nc-pot}
  k\frac{\partial U}{\partial k}
  &= -\frac{N_f N_c}{8\pi^2} \; \frac{k^6}{k^2\!+\!M_q^2} \; , \\  
\label{eq:large-nc-z}
  k\frac{\partial Z_\Phi}{\partial k} &= -\frac{N_f N_c}{4\pi^2} \,
  \frac{k^6}{(k^2\!+\!M_q^2)^3} \; .
\end{align}
The right hand sides of these equations only depend on the quark mass Eq.
(\ref{eq:masses}), but not on the potential $U$ or the wave function
renormalization parameter $Z_\Phi$, therefore they completely decouple and can
be integrated analytically. In particular, these equations do not
contain a RG improvement. Note that in the large $N_c$ limit we have
$M_q^2 = \Phi^2$ and the integrals read
\begin{align}
  \int \limits^{U(k)}_{U_{\mathrm{uv}}} \; d U  &=  -\frac{N_f
  N_c}{8\pi^2} \, \int \limits_{k_{\mathrm{uv}}}^{k}
  \frac{\kappa^5}{\kappa^2\!+\!\Phi^2} d \kappa \; ,\\ 
  \int \limits^{Z_\Phi(k)}_{Z_{\Phi}^{\mathrm{uv}}} dZ_\Phi & =
  -\frac{N_f N_c}{4\pi^2} \,
  \int \limits_{k_{\mathrm{uv}}}^{k} \,
  \frac{\kappa^5}{(\kappa^2\!+\!\Phi^2)^3} d \kappa \; .
\end{align}
Neglecting $\Phi$-independent terms, the integral for 
the potential $U$ yields
\begin{align}
\label{eq:solution-potential}
  U(\Phi,k) = U_{\mathrm{uv}}(\Phi) +\frac{N_f N_c}{16\pi^2} \,\left(
  \Phi^4 \log \left( \frac{k_{\mathrm{uv}}^2 + \Phi^2}{k^2 + \Phi^2}
  \right) -\Phi^2 (k_{\mathrm{uv}}^2-k^2) \right) \; ,
\end{align}
where the UV potential is given by Eq. (\ref{eq:uv-potential}). Using
$Z_{\Phi}^{\mathrm{uv}}\!=\!0$ as starting value, the equation for
$Z_\Phi$
can be integrated in the same way giving a slightly longer
expression
\begin{align}
\label{eq:solution-zphi}
  Z_\Phi(\Phi,k)&=\frac{3}{8{\pi }^2} \left( 2\,\log
           \left(\frac{k_{\mathrm{uv}}^2 + {\Phi }^2} {k^2 + {\Phi
           }^2}\right) \right. \nonumber\\ 
          &\left. \quad \qquad -\frac{\left(
           k_{\mathrm{uv}}^2-k^2 \right) \, {\Phi }^2\,\left(
           4\,k^2\,k_{\mathrm{uv}}^2 +  3\,\left( k^2 + k_{\mathrm{uv}}^2
           \right) \, {\Phi }^2 + 2\,{\Phi }^4 \right) }{
         {\left( k^2 + {\Phi }^2 \right) }^2\,
         {\left( k_{\mathrm{uv}}^2 + {\Phi }^2 \right) }^2} \right) \; .
\end{align}
Eqs. (\ref{eq:solution-potential}, \ref{eq:solution-zphi}) are the
exact analytic solution of
the large $N_c$ flow equations for a given scale $k$. To obtain the
physical vacuum expectation value $\Phi_0(k)$ of the field, the
potential has to be minimized with respect to $\Phi$:
\begin{equation}
\label{eq:rgvac}
  \left. \frac{\delta U(\Phi,k)}{\delta \Phi^2}
  \right|_{\Phi=\Phi_0} =\frac{m^2}{2}+\frac{N_f N_c}{16\pi^2}
  \left( 2\Phi_0^2 \log \left( \frac{k_{\mathrm{uv}}^2+\Phi_0^2}{\Phi_0^2}
  \right) -
  \frac{k_{\mathrm{uv}}^2(k_{\mathrm{uv}}^2+2\Phi_0^2)}{k_{\mathrm{uv}}^2+
  \Phi_0^2} \right) =0 \; .
\end{equation}
This equation cannot be solved analytically due to the non algebraic
logarithmic term. As we will show, it corresponds to the self
consistent NJL gap equation at the IR scale $k\!=\!0$, determining the
physical quark mass $m_q = \Phi_0(k \!=\! 0)$ as the minimum of the
potential Eq. (\ref{eq:solution-potential}). 
The full potential provides more information
than the NJL gap equation, since it gives arbitrary higher order
moments of $\Phi$ besides its vacuum expectation value. This allows to
compute for instance the spectrum of the Dirac operator which will be
presented in a forthcoming paper \cite{diracspectrum}. \\
The NJL model in large $N_c$ approximation is standardly considered in
a self consistent approach which leads to the NJL gap equation
\cite{Ripka}. This equation determines the constituent quark
mass and is usually regularized by a hard cutoff. Furthermore, to make
contact with chiral degrees of freedom there is an additional
equation that determines the mesonic wave function renormalization
factor $Z_\Phi$, once the constituent quark mass is known \cite{Ripka}. \\
To compare with our flow equations we regularize these equations in a
Schwinger proper time scheme using the heat kernel cutoff. Since the
heat kernel cutoff $f^{(n)}(k^2\tau)$ acts like a IR cutoff
function, the desired UV regulation is given by the function $(1 -
f^{(n)}(k_{\mathrm{uv}}^2\tau))$, which cuts off momenta larger than
$k_{\mathrm{uv}}$. The gap equation and the determining equation
for $Z_\Phi$ read
\begin{align}
\label{eq:gapeq}
1 &= 8  G N_c N_f  \int_0^\infty d\tau \, \int \! \frac{d^4 p}{(2
\pi)^4} \;\; e^{-\tau(p^2+m_q^2)}(1 - f^{(n)}(k_{\mathrm{uv}}^2\tau)) \;
, \\
\label{eq:zphieq}
Z_\Phi &= 2 N_c N_f \int_0^\infty \tau d\tau \, \int \! \frac{d^4 p}{(2
\pi)^4} \;\; e^{-\tau(p^2+m_q^2)}(1 - f^{(n)}(k_{\mathrm{uv}}^2\tau)) \;
.
\end{align}
The integrals appearing in these equations can be integrated
analytically. With Eq. (\ref{eq:gnjl}) the NJL coupling $G$
is connected to the UV meson mass $m_{\mathrm{uv}}^2$. For the cutoff $f^{(2)}$
the integrated form of the gap equation Eq. (\ref{eq:gapeq}) is
analytically equal to Eq. (\ref{eq:rgvac}). Similarly, the determining
equation for $Z_\Phi$ Eq. (\ref{eq:zphieq}) yields
Eq. (\ref{eq:solution-zphi}) evaluated at $\Phi^2\!=\!\Phi_0^2$ after
integration. This connection between the equations in the RG scheme and
the equations in the self consistent NJL approach holds for all cutoff
functions $f^{(n)}$ Eq. (\ref{eq:co}), provided both schemes are
regularized with the same cutoff. This shows that the RG flow result
is {\em exactly identical} to the standard NJL approach in large $N_c$
approximation. \\
Although the vacuum has to be computed numerically, there are two
quantities, that allow an analytic solution. The first one is the
scale $k_{\chi SB}$, where the system evolving from the UV enters the
chiral broken regime, which is given as the solution of the equation
\begin{equation}
  \left. \frac{\delta U(\Phi,k)}{\delta \Phi^2} \right|_{\Phi^2=0}=0
  \; ,
\end{equation}
and takes the simple form
\begin{equation}
\label{eq:kxsb}
  k_{\chi \mathrm{SB}} = \sqrt{k_{\mathrm{uv}}^2 - \frac{8 \pi^2}{N_c N_f}
  m^2_{\mathrm{uv}}} \; .
\end{equation}
The other one is the ratio $m_\sigma/m_q$, which results from
Eqs. (\ref{eq:masses}, \ref{eq:physmasses}) using the analytic expressions
for $U$ and $Z_\Phi$ at $k=0$ and $\Phi=\Phi_0$. It is exactly 2 in large
$N_c$, as known from the standard NJL approach \cite{Ripka}. \\
For a  numerical solution, we adjust the UV parameters to reproduce
physical values of $m_q\!=\!320$ MeV and $f_\pi\!=\!93$ MeV in the
case of an explicit chiral symmetry breaking, discussed in the next
chapter. The initial parameters for the RG evolution read
\begin{align}
\label{eq:initial-value-NJL-m} 
\begin{split}
k_{\mathrm{uv}} &= 1037 \; \mbox{MeV} \; , \\
m_{\mathrm{uv}} &= 228.0 \; \mbox{MeV} \; , \\
Z_\Phi^{\mathrm{uv}}  &= 0 \; ,
\end{split}
\end{align}
and $Z_q=g=1=$ const. This corresponds to a four fermion coupling of
$G=9.62$ GeV$^{-2}$.
With these initial values we calculate the
effective potential from Eq. (\ref{eq:solution-potential}) and the
wave function renormalization from the corresponding equation for
$Z_\Phi$. The result is the renormalized effective potential $U_R$,
which is shown in Fig. \ref{fig:pot-plot} as a function of the field
$\Phi_R$ and the scale $k$. From this potential we 
compute the scale dependent pion decay constant $f_\pi(k)$ shown as
white curve in Fig. \ref{fig:pot-plot}.
 \\
The quark mass and meson
masses, which are given from Eq. (\ref{eq:masses}) are plotted in Fig.
\ref{fig:masses} (solid lines).
In the chiral symmetric regime for
scales $k\!>\!k_{\chi SB}\!=\!626$ MeV (cf. Eq. (\ref{eq:kxsb})) both meson
masses are identical, while the quark mass is zero. 
Below the symmetry breaking scale the pions become
massless while the $\sigma$-meson mass rises again. Due to the
nontrivial vacuum of the effective potential, the quarks acquire a
finite mass. In the IR limit $k=0$ we find the physical values
for the the quark mass $m_q \!=\! \Phi_0(k\!=\!0)$ and the $\sigma$-meson
mass Eqs. (\ref{eq:masses},\ref{eq:physmasses}). \\
The Yukawa coupling is
given by $g_R \!=\! Z_\Phi^{-1/2}(k\!=\!0)$ and the pion decay constant is
$f_\pi\!=\!Z_\Phi^{1/2}(k\!=\!0) \, \Phi_0(k\!=\!0)$. The IR values
are summarized in Table \ref{tab:values} (second column).
\begin{table}[h]
  \begin{center}
    \begin{tabular}{lr||c|c|c||}
     & & $m_c\!=\!7$MeV & $\chi$-limit & $\Phi^4$-truncation \\
     \hline \hline
     $f_\pi$ &[MeV]                  & 93.0 & 92.0 & 105 \\
     $\lambda$ &                     & 23.6 & 22.6 & 14.9 \\
     $g_R\!=\!Z_\Phi^{-1/2}(\Phi_0^2) $ &      & 3.44 & 3.36 & 2.73 \\
     $m_q\!=\!\Phi_0$ &[MeV]         & 320 & 309 & 285 \\
     $m_\sigma$ &[MeV]               & 650 & 618 & 570 \\
     $m_\pi$ &[MeV]                  & 116 & 0 & 0 
    \end{tabular}
\vspace{0.2cm}
    \caption{The IR values for different approximations are shown. All
     computations are done with the UV initial conditions
      $k_{\mathrm{uv}}\!=\!1037$ MeV, $m_{\mathrm{uv}}\!=\!228.0$ MeV and
      $Z_{\Phi}^{\mathrm {uv}}\!=\!0$. The first column shows the analytic RG
      result with explicit chiral symmetry breaking due to a finite
      current quark mass of $7$ MeV. In the second column the analytic RG
      result in the chiral limit is given, which is equal to the
      result of the NJL gap-equation regularized with the proper time
      regulator. In the last column the results of a simple quartic truncation
      of the potential are shown.
      }
    \label{tab:values}
  \end{center}
\end{table}

\section{Explicit symmetry breaking}
Although the physical number of three colors is not really large, we
know from standard NJL computations, that the large $N_c$ limit gives a
reasonable approximation. We next compare our model with the physical
case of an explicit chiral symmetry breaking due to a finite current
quark mass $m_c$. \\
The value of the mean current quark mass $m_c\!=\!\frac{m_u+m_d}{2}\!=\!7$ MeV
is taken from chiral perturbation theory \cite{Gasser}, which has been
performed at our UV scale of 1 GeV, although there is some uncertainty
because of the different schemes. From Eq. (\ref{eq:gnjl}) the
parameter $\delta$ can be computed resulting in $\delta\!=\!3.64 \cdot
10^{-4} (\mathrm{GeV})^3$. As already discussed, the solution Eq.
(\ref{eq:solution-potential}) obtains in the case of explicit symmetry
breaking, with a term linear in $\sigma$ in the UV potential Eq.
(\ref{eq:uv-potential}). The vacuum expectation value is now given by
minimization with respect to $\sigma$. The evolution of the
renormalized physical masses is shown as dashed curves in Fig.
\ref{fig:masses}, which exhibits a very reasonable splitting of the
mass scales. There is a continuous transition in the $k$ evolution
from the current quark to the constituent quark regime, since the
system is in a nontrivial vacuum right from the beginning of the
evolution. The quark mass assumes its current quark mass in the UV and
the pion looses its role as an exact Goldstone boson and
acquires a small finite mass in the IR. \\
The IR values of the physical quantities are shown in the first
column of Table \ref{tab:values}, The effect of the explicit chiral
symmetry breaking is a minor change in the other parameters and a finite
pion mass of $116$ MeV which is reasonable close to the physical
value. The change of the pion decay constant compared to the chiral
limit amounts only to 1 MeV and is rather small in large $N_c$
approximation compared to chiral perturbation theory
computations \cite{Gasser}. However, this seems to change at order
$1/N_c$ \cite{fullflow}.

\section{Truncated potential}
\begin{figure}[h]
\begin{center}
\begin{raggedright}
\epsfig{file=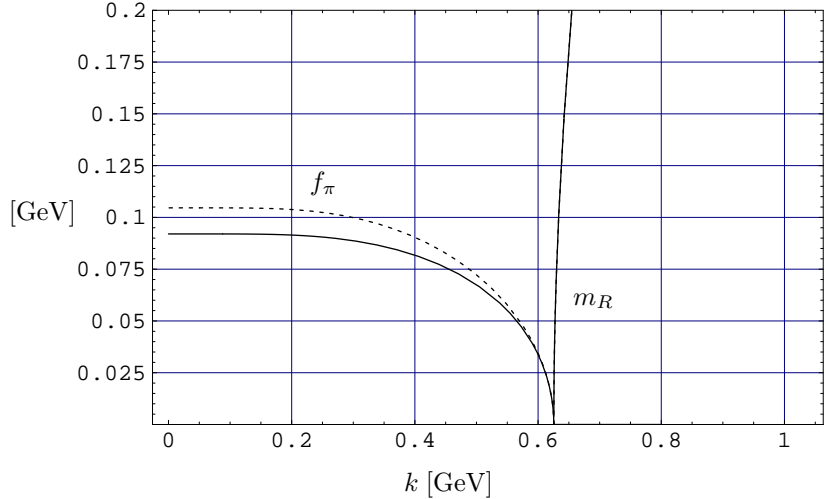,width=10cm}
\end{raggedright}
\centerline{$k \; [\mbox{GeV}]$}
\flushleft \vspace{-4.7cm} \hspace{4.1cm} {$f_\pi$}
\flushleft \vspace{-0.3cm} \hspace{0.1cm} {$[\mbox{GeV}]$}

\vspace{0.7cm} \hspace{7.6cm} {$m_R$}
\vspace{+2.5cm} 
\caption{\label{fig:m-fpi}The renormalized meson mass
  $m_R\!=\!m_\sigma\!=\!m_\pi$ in the symmetric regime and the pion
  decay constant in the broken regime as functions of $k$ for the
  exact solution (solid) and the quartic truncation (dotted). In the
  symmetric regime the two solutions are equal, but the vacuum results
  for $f_\pi$ differ by $14\%$.
}
\end{center}
\end{figure}

\begin{figure}[h]
\begin{center}
\begin{raggedright}
\epsfig{file=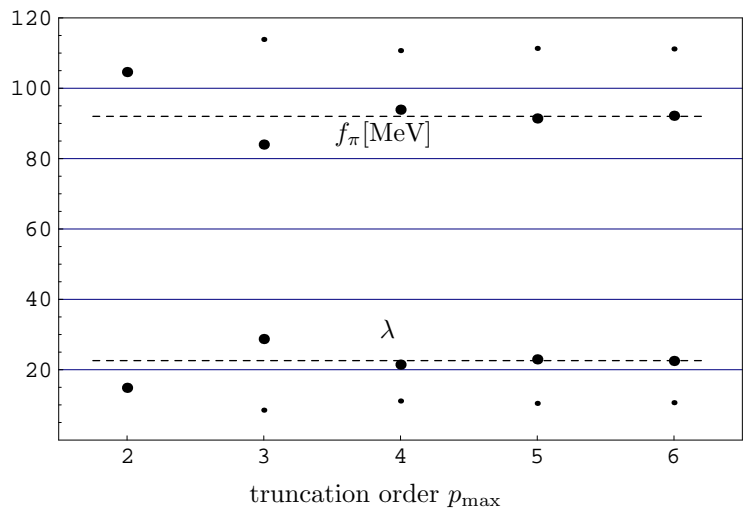,width=10cm}
\end{raggedright} 
\centerline{truncation order $p_{\mathrm{max}}$}
\flushleft \vspace{-5.5cm} \hspace{5.4cm} {$f_\pi[\mbox{MeV}]$}

\vspace{2.2cm} \hspace{6cm} {$\lambda$}
\vspace{+2.3cm} 
\caption{\label{fig:convergence} The dependence of the physical pion
  decay constant $f_\pi$ and the 4-meson coupling $\lambda$ on the
  truncation order $p_{\mathrm{max}}$ of a polynomial
  parametrization. Thin points give the behaviour when only the
  potential $U$ is expanded and the wave function renormalization
  factor $Z_\Phi$ is considered $\Phi$-independent. They do not
  converge to the exact results given by the dashed lines. Thick
  points show the dependence when $U$ and $Z_\Phi$ are expanded and
  exhibit a fast convergence to the exact results.}
\end{center}
\end{figure}

The analytic solution of Eqs.
(\ref{eq:large-nc-pot},\ref{eq:large-nc-z}) can test the usual
parametrizations of the model. One relies on such
parametrizations, if higher corrections are regarded, e.g. mesons are
taken into account \cite{RGreview,Schaefer:1999em}. A standard
approach is to expand the model in powers of
$\Phi^2$ up to a given order $p_{\mathrm{max}}$. \\
The potential $U$ evolves under the renormalization flow from a
trivial to a nontrivial form. Therefore we consider two
parametrizations. In the simplest quartic truncation with
$p_{\mathrm{max}}\!=\!2$, they read in the chiral symmetric regime
$U(\Phi ^2,k)=\frac{m^2(k)}{2} \Phi^2+\frac{\lambda(k)}{4}(\Phi ^2)^2$
and in the broken regime $U(\Phi ^2,k)=\frac{\lambda(k)}{4} (\Phi
^2\!-\! \Phi_0 ^2(k))^2$.  Since the wave function renormalization
factor $Z_\Phi$ arises together with the
dynamical term of mass dimension four, it is correspondingly expanded
up to order $p_{\mathrm{max}}\!-\!2$ and is therefore
$\Phi$-independent in the simplest approximation. \\
The evolution equations for the couplings $m^2$ and $\lambda$ 
can be derived from Eq. (\ref{eq:large-nc-pot}). We find for
large $N_c$ in the symmetric case ($g\!=\!Z_q\!=\!1$):
\begin{align}
\label{eq:large-nc-m}
k \frac{\partial m^2}{\partial k} & = 
+ \frac{N_cN_f}{4 \pi^2} \,k^2 \; , \\
k \frac{\partial \lambda}{\partial k} & = 
- \frac{N_cN_f}{2\pi^2} \; , \\
k\frac{\partial Z_{\Phi}}{\partial k} & = 
- \frac{N_cN_f}{4 \pi^2} \; ,
\end{align}
and for the broken regime:
\begin{align}
\label{eq:large-nc-phi-broken}
k\frac{\partial \Phi^2_{0}}{\partial k} & = 
- \frac{N_cN_f}{4\pi^2} \, \frac{1}{\lambda} \,
\frac{k^2}{\left(1+\Phi^2_{0} / k^2\right)^2} \; , \\
k \frac{\partial \lambda}{\partial k} & = 
- \frac{N_cN_f}{2\pi^2} \,
\frac{1}{\left(1+\Phi^2_{0} / k^2\right)^3} \; , \\
k \frac{\partial Z_{\Phi}}{\partial k} & = 
- \frac{N_cN_f}{4 \pi^2} \,
\frac{1}{\left(1+\Phi^2_{0} / k^2\right)^3}\; .
\label{eq:large-nc-Z-broken}
\end{align}
To solve these equations numerically, we use again the initial
values Eq. (\ref{eq:initial-value-NJL-m}). In addition we choose
$\lambda_{\mathrm{uv}}\!=\!0$ to have the same UV potential as
before (cf. Eq. \ref{eq:uv-potential}).
In the symmetric regime the exact and approximate solutions are
identical.  In the broken regime we switch to the second
parametrization and follow the flow given by Eqs.
(\ref{eq:large-nc-phi-broken}-\ref{eq:large-nc-Z-broken}) to $k=0$.
Now we have a set of {\em coupled \/} equations.  This comes from the
fact that we are tracking the evolution of the minimum $\Phi_0$ of the
potential $U$, which has no analytical solution from Eq.
(\ref{eq:solution-potential}). In the IR we find a pion decay
constant $f_\pi\!=\!105$ MeV compared to a
value of 92 MeV for the exact solution (cf. Fig. \ref{fig:m-fpi}). The
IR masses and couplings from these truncated flow equations are 
given in the third column of Table \ref{tab:values}.
\newpage
When the truncation is extended to higher orders $p_{\mathrm{max}}$, the IR
value of $f_\pi$ converges fast to the exact solution with a pion
decay constant of 92 MeV, as indicated by the thick points in
Fig. \ref{fig:convergence}. On the other hand, if only the effective
potential is expanded, the flow still converges, however to a strongly
deviating result of 111 MeV, shown by the thin points in
Fig. \ref{fig:convergence}. This is in contrast to a naive dimensional
analysis, which would suggest that the higher order momenta of $Z_\Phi$
should be suppressed by their high mass dimension in the vicinity of
an IR fixed point. Although this does not seem to be the case in large
$N_c$ it may be the case when meson corrections are included, where a
partial fixed point behaviour occurs. 
Nevertheless, in order to obtain a credible solution it is
necessary to choose the truncation order in such away, that convergence with
respect to {\em all \/} model parameters is achieved. Indeed, at order
$p_{\mathrm{max}}\!=\!7$ the results are to the given accuracy exactly
the ones of the exact solution in the second column of Table
\ref{tab:values}.

\section{Conclusion}
We have derived RG flow equations for the linear $\sigma$-model in
large $N_c$ approximation as a different approach to solve the NJL
model. It was shown that in large $N_c$ the RG approach reproduces the
NJL gap equation and therefore in this limit the solutions of the two
schemes are equal. In
addition one finds from the flow equations the whole effective
potential, instead of just its minimum. 
In the case of explicit symmetry breaking due to a finite
current quark mass we find very reasonable meson- and constituent
quark masses.  A comparison of the full flow equations with a
polynomial truncation has shown a deviation of $\approx 15\%$ in
lowest order, but a fast convergence to the full result. \\
Both, the exact and truncated calculation use two parameters
$k_{\mathrm{uv}}$ and $m_{\mathrm{uv}}$ to fix the quark mass $m_q$ and
the pion decay constant $f_{\pi}$. This is the conventional method in
NJL theory. An advantage of RG equations is the reduction of the
relevant UV parameters, if the theory exhibits a partial IR fixed point
behaviour. Such a fixed point behaviour indeed occurs when meson loops
are considered.\\
In addition, meson loop corrections can be included systematically
without the problem of introducing additional cutoffs, since the
effective action is renormalizable. The analytic large $N_c$ solution
presented here serves as a reference to the analysis of the full flow
equations with meson fluctuations, which will be published in a
forthcoming paper \cite{fullflow}.

\section*{Acknowledgments}
J.~M. and K.~S. would like to thank O.~Bohr, B.-J.~Schaefer and T.~Spitzenberg
for helpful discussions.


\newpage 
\begin{thebibliography}{99}
\bibitem{Meggiolaro:2000kp}
E.~Meggiolaro and C.~Wetterich,
Nucl. Phys. B {\bf 606} (2001) 337,
hep-ph/0012081.

\bibitem{Klevansky:1992qe}
S.~P.~Klevansky,
Rev.\ Mod.\ Phys.\  {\bf 64} 649 (1992);
T.~Hatsuda and T.~Kunihiro,
Phys.\ Rept.\  {\bf 247} (1994) 221,
hep-ph/9401310.

\bibitem{Oertel:2000fk}
M.~Oertel, M.~Buballa and J.~Wambach,
Phys.\ Lett.\ B {\bf 477} (2000) 77,
hep-ph/9908475;
M.~Oertel, M.~Buballa and J.~Wambach,
Nucl.\ Phys.\ A {\bf 676} (2000) 247,
hep-ph/0001239,
G.~Ripka,
Nucl.\ Phys.\ A {\bf 683} (2001) 463,
hep-ph/0003201.

\bibitem{RGbasics}
K.~G.~Wilson and J.~Kogut, Phys. Rep. {\bf 12} (1974) 75;
F.~J.~Wegner and A.~Houghton, Phys.\ Rev.\ A {\bf 8} (1973) 401;
C.~Wetterich, Phys. Lett. B {\bf 301} (1993) 90;
M.~Bonini, M. D'Attanasio and G. Marchesini, Phys. Lett. B {\bf 409}
(1993) 441;
T.~R.~Morris, Int. J. Mod. Phys. A {\bf 9} (1994) 2411.


\bibitem{RGreview}
J.~Berges, N.~Tetradis and C.~Wetterich,
hep-ph/0005122;
J.~Berges, 
Lectures given at 11th Summer School 
and Symposium on Nuclear Physics (NuSS 98), hep-ph/9902419.


\bibitem{Schaefer:1999em}
B.~Schaefer and H.-J.~Pirner, Nucl. Phys. A {\bf 660} (1999) 439,
nucl-th/9903003;
B.-J.~Schaefer and H.-J.~Pirner,
Nucl.\ Phys.\ A {\bf 627} (1997) 481
hep-ph/9706258;
J.~Meyer, G.~Papp, H.-J.~Pirner and T.~Kunihiro,
Phys.\ Rev.\ C {\bf 61} (2000) 035202,
nucl-th/9908019;
G.~Papp, B.-J.~Schaefer, H.-J.~Pirner and J.~Wambach,
Phys.\ Rev.\ D {\bf 61}, 096002 (2000),
hep-ph/9909246.

\bibitem{Litim:2001ky}
D.~F.~Litim and J.~M.~Pawlowski,
hep-th/0111191.

\bibitem{Ripka}
G.~Ripka, ``Quarks bound by chiral fields'', Oxford University Press
(1997).

\bibitem{Zinn-Justin}
J.~Zinn-Justin, ``Quantum field theory and critical phenomena'',
Oxford 1989. 

\bibitem{Litim}
D.~F.~Litim and J.~M.~Pawlowski,
Phys.\ Lett.\ B {\bf 516} (2001) 197,
hep-th/0107020.

\bibitem{fullflow}
J.~Meyer, K.~Schwenzer and H.-J.~Pirner, in preparation.

\bibitem{Bohr}
O.~Bohr, B.-J.~Schaefer and J.~Wambach,
Int.\ J.\ Mod.\ Phys.\ A {\bf 16} (2001) 3823,
hep-ph/0007098.

\bibitem{diracspectrum}
T.~Spitzenberg, K.~Schwenzer and H.-J.~Pirner, in preparation




\bibitem{Gasser}
J.~Gasser and H.~Leutwyler,
Phys.\ Rept.\  {\bf 87} (1982) 77.


\end{thebibliography}
\end{document}